# Role of Centrosymmetry in the Photophysics of Molecular Aggregates

Qingyun Wan[1]* and Chi-Ming Che[1, 2, 3]*

[1] Department of Chemistry and State Key Laboratory of Synthetic Chemistry, The University of Hong Kong, Pokfulam Road, Hong Kong, China.

[2] HKU Shenzhen Institute of Research & Innovation, Shenzhen 518057, China.

[3] Hong Kong Quantum AI Lab Limited Units 909-915, Building 17W, 17 Science Park West Avenue, Pak Shek Kok, Hong Kong, China.

To understand the photophysics of molecular aggregates, exciton model of J- and H-aggregate has been extensively utilized. However, it lacks consideration of crystal symmetry. Although discrete molecules may lack symmetry, their aggregates can exhibit a high degree of symmetry. Herein, we utilized group theory to study the optical properties of centrosymmetric molecular aggregates, showing that their optical selection rules (transition dipole moment and spin-orbit coupling) are determined by the symmetry of singlet and triplet excited states and the intermolecular orbital overlap. Symmetry-forbidden electronic transitions are closely related to ultralong organic phosphorescence. Our model's scope is broad, as over 50% of organic crystals belong to centrosymmetric space groups according to Cambridge Structural Database.

The photophysics of molecular aggregates have been a hot area in the field of soft matter for decades, as molecular aggregates display unique excited state properties different from discrete molecules, such as super-radiance of J-aggregate, ultralong organic phosphorescence (UOP), and aggregation-induced emission (AIE) [1-3]. The molecular exciton model, developed by Kasha and co-workers in 1965[1], has been widely used to understand photo-physical properties of solid-state molecules. This model demonstrates how electrostatic dipole–dipole interactions between neighboring molecules affect the exciton properties of aggregates (FIG. 1a) [1]. Subsequently, Hoffmann, Spano, and others extended the molecular exciton model to account for the effects of molecular vibrations and intermolecular charge transfer on the photophysics of molecular aggregates [1,4-8]. A comprehensive review by Spano and coworkers, building on Kasha's work, summarizes the development of molecular exciton models [4]. Kasha's exciton model explains the influence of molecular aggregation on radiative decay processes. Regarding non-radiative processes, molecular aggregation can restrict intramolecular rotation, resulting in a slower non-radiative decay and contributing to the AIE phenomenon [3,9,10], which has attracted significant attention in recent studies (Fig. 1b) [11].

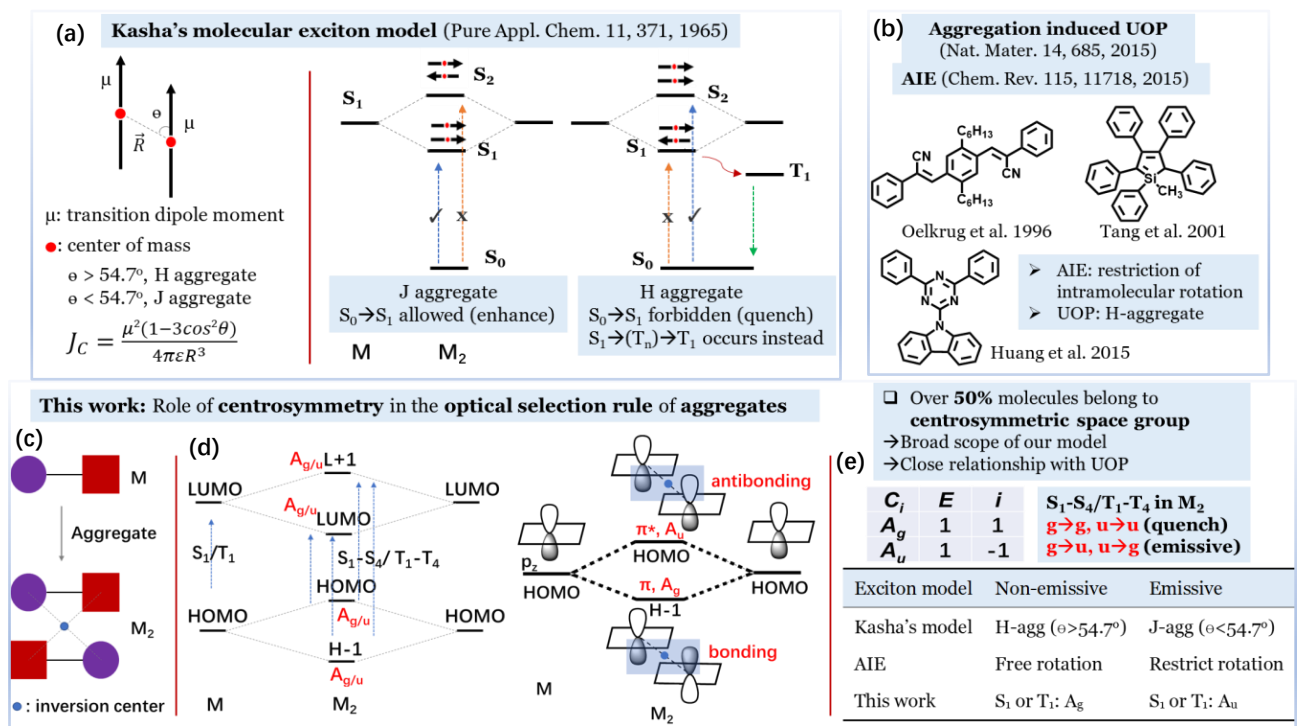

FIG. 1. (a) Illustration schemes for Kasha's molecular exciton model. $\varepsilon$ is the dielectric constant of the medium. $J_C$ is the Coulomb coupling between two molecules. μ is the transition dipole moment of monomer at $S_1$ state. (b) Important emitters related to AIE and UOP phenomena. (c) Illustration schemes for the aggregation process. (d) A simplified scheme showing the formation of $A_g$ and $A_u$ orbitals between two monomers by intermolecular orbital overlap. M and $M_2$ indicate the monomer and dimer respectively. g or u indicate even or odd under inversion. (e) Characteristic table of point group $C_i$ and the selection rules, together with a summary of all molecular exciton models.

Although Kasha's exciton model and AIE related research have been extensively studied and developed, these models lack proper consideration of crystal symmetries in the optical selection rules when applied to solid-state molecules. The optical selection rule is closely related to a molecule's symmetry. Take Laporte rule as an example, d-d electronic transitions of octahedral organometallic complexes with a center symmetry are symmetry-forbidden, leading to the low absorptivities [12]. For small molecules like benzene and naphthalene, the relationship between their crystal symmetries and optical transitions has been established previously [13]. However, nowadays, the importance of crystal symmetry in the photophysics of molecules in aggregated/crystalline form has seldom been addressed. In this study, we studied the molecular exciton in aggregates based on crystal symmetry and group theory considerations, combined with TDDFT (time-dependent density functional theory) and MO (molecular orbital) calculations. Our objective is to determine the allowance of electronic transitions in molecular aggregates, namely $S_1 \rightarrow S_0$ (fluorescence) and $T_1 \rightarrow S_0$ (phosphorescence) transitions ($S_1/T_1$: first singlet/triplet excited state, $S_0$: ground singlet state). We find that both $S_1 \rightarrow S_0$ and $T_1 \rightarrow S_0$ transitions in the molecular aggregates are symmetry-forbidden when $S_1$ and $T_1$ states are totally symmetric. The orbital overlap and formation of bonding and antibonding MOs between two adjacent monomers play a key role in determining the symmetry of the excited state of the aggregate. The observation that organic molecules preferentially crystallize in centrosymmetric space groups, with over 50% belong to two centrosymmetric groups of $P2_1/c$ and P ī (2) [14], suggests the broad scope of our exciton model with the symmetry considerations.

We further demonstrated a close relationship between the centrosymmetry and UOP in molecular aggregates. UOP describes that pure organic molecular aggregates without heavy metal atom could exhibit emission in the second time regime at room temperature. Organic crystals with UOP have diverse potential applications, such as in optoelectronics, photodynamic therapy, nonlinear optics, photochemical upconversion and bioimaging, and have received a surge of interest in recent years [2,15]. In this study, we checked the packing forms of 27 molecular crystals (Scheme S1) that were reported to display UOP. We found that 25 molecules (**1-2** in FIG. 2a, **S1-S5**, **S7-S14**, **S16-S25** in SCHEME S1), which belong to the space group of Pnma (62), C2/c (15), Pbca (61), $P2_1/n$ (14) or Pī (2), all have inversion symmetry in their respective crystal structures (TABLE SI). By our calculations, 18 of these 25 molecules have a symmetry forbidden $S_1 \rightarrow S_0$ and $T_1 \rightarrow S_0$ transition, resulting in UOP with a small radiative decay rate and long-lived emission lifetime.

As two examples, we performed detailed calculations on dimer **1**[16] and **2**[17] (FIG. 2a). Compound **1** and **2** were reported to show UOP at crystalline state with the lifetime of 12.7 ms and 120 ms in literatures, respectively [16,17]. As depicted in FIG. 2b and FIG. S2, TDDFT calculations show that in dimer $[\mathbf{1}]_2$ and $[\mathbf{2}]_2$, $f(S_1)$ (oscillator strength of $S_1$) equals 0. While for monomer **1** and monomer **2**, a relatively large $f(S_1)$ of 0.7 and a moderate $f(S_1)$ of 0.02 were calculated respectively. To understand these forbidden electronic transitions of the dimer, we analyzed the symmetry of the MOs involved in the $S_1 \rightarrow S_0$ transitions of $[\mathbf{1}]_2$ and $[\mathbf{2}]_2$. As shown in FIG. 2b. The HOMO of **1** is mainly located at the carbazole group, and the LUMO at the Bodipy group.

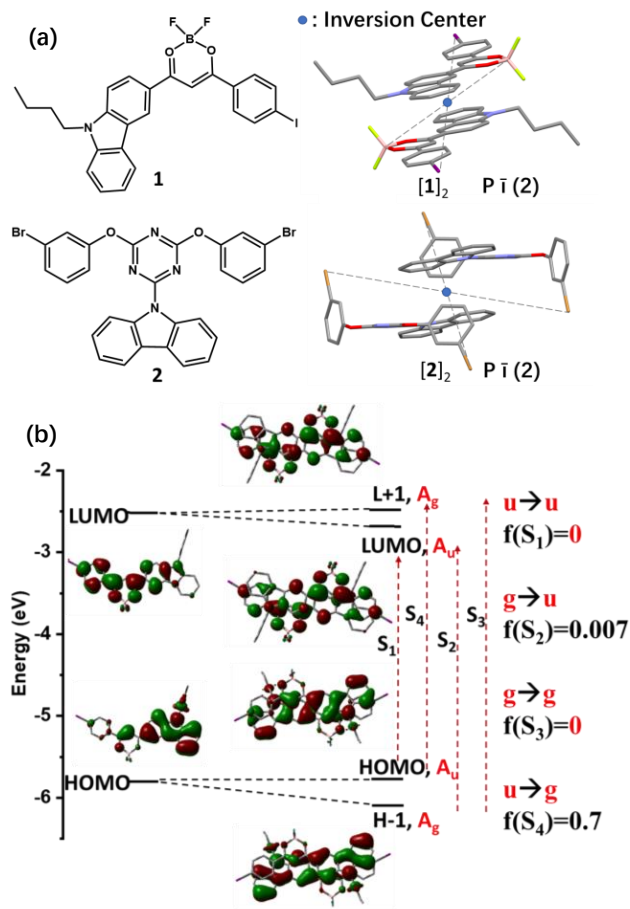

FIG. 2. (a) Chemical (left) and crystal packing (right) structures of compound **1** and **2**. (b) Calculated molecular orbital (MO) diagram for **1**, $[\mathbf{1}]_2$, together with the calculated $f(S_1)$ for monomer, and $f(S_1-S_4)$ for dimer. For clarity, only major transitions with contributions over 50% in $S_1-S_4$ states are shown here.

When two **1** monomers are in proximity, the HOMO and LUMO orbitals of the monomer overlap, leading to the formation of four orbitals in dimer $[\mathbf{1}]_2$, i.e., H-1, HOMO, LUMO and L+1 orbitals. The symmetry of H-1, HOMO, LUMO and L+1 in $[\mathbf{1}]_2$ is identified as $A_g$, $A_u$, $A_u$ and $A_g$, respectively, under the point group of $C_i$. FIG. 1e gives the character table for point group $C_i$, together with the selection rule. The value of $f(S_1)$ depends on the $<S_1|\mu_e|S_0>$ bra-ket and is nonzero (symmetry allowed) when the direct product of $\Gamma(S_1) \times \Gamma(\mu_e) \times \Gamma(S_0)$ is or contains a totally symmetric representation of $A_g$. $\mu_e$ is the electric dipole operator. Based on the optical selection rule constrained by centrosymmetry, $S_1$ (u→u) and $S_3$ (g→g) transitions in $[\mathbf{1}]_2$ are forbidden, while $S_2$ (g→u) and $S_4$ (u→g) transitions in $[\mathbf{1}]_2$ are allowed (FIG. 2b). An examination of the symmetry of MOs involved in the electronic transitions of $[\mathbf{2}]_2$ leads to the conclusion that the $S_1$ (u→u) and $S_2$ (u→u) transitions in $[\mathbf{2}]_2$ are symmetry forbidden, while the $S_3$ (u→g) and $S_4$ (u→g) transitions in $[\mathbf{2}]_2$ are symmetry allowed (FIG. S2).

We further use natural transition orbitals (NTOs) to analyze the TDDFT excitations of $[1]_2$. The NTO results shown in FIG. S3 reveal that the $S_1$ (u→u) and $S_3$ (g→g) transitions are symmetry-forbidden, $S_2$ (g→u) and $S_4$ (u→g) transitions are allowed for $[1]_2$, consistent with the MO analysis.

The orbital overlap between the two monomers plays an important role in determining the symmetry ($A_g$ or $A_u$) of the MOs in the dimer. Intermolecular π-π interactions are considered here, which lead to the formation of bonding π and antibonding π* orbitals between the two monomers. As shown in FIG. 1d, the π orbital is centrosymmetric, which belongs to the $A_g$ subgroup, while the antibonding π* orbital undergoes inversion symmetry breaking and belongs to the $A_u$ subgroup. Longer-range interactions in aggregated networks of $[1]_n$ (n =1-10) are considered. In extended molecular aggregates of $[1]_n$, the population of excitons on even (n = 4, 6…) or odd numbers (n = 3, 5, 7…) of molecules in the aggregate affects the optical selection rule. As shown in FIG. 3, aggregates $[1]_n$ (n = 2, 4, 6, 10) have $A_g$ symmetry in the $S_1$ state, leading to strictly forbidden electronic transitions with $f(S_1)$ equals 0. The $A_u$ symmetry of $S_1$ is calculated for $[1]_8$, leading to a slightly allowed $S_1$ state. For aggregates containing an odd number of molecules, the optical selection rule relaxes due to the breaking of inversion symmetry. The value of $f(S_1)$ decreases with increasing chain length containing an odd number of molecules. In **1**, $f(S_1)$ is 0.7 and decreases to 0.0053 in $[1]_3$, further to 0.0001 in $[1]_7$ and $[1]_9$. Examination of excited state's symmetry in three dimensional (3D) molecular aggregates was also performed, revealing forbidden $S_1$ transitions in a 3D $[1]_4$ tetramer and intermolecular orbital interactions from other dimensions (FIG. S4).

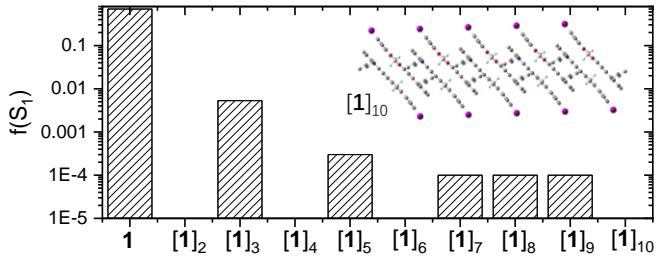

FIG. 3. Calculated $f(S_1)$ for one dimensional $[1]_n$ (n=1-10).

TABLE I. Calculated $f(S_1)$ and symmetry of $S_1$, H-1, HOMO, LUMO, and L+1 for dimer $[1]_2$ at different $\theta$.

| θ(°) | H-1 | HOMO | LUMO | L+1 | Transitions for $S_1$ | $S_1$ | $f(S_1)$ |
|---|---|---|---|---|---|---|---|
| 22 | $A_g$ | $A_g$ | $A_g$ | $A_u$ | HOMO→LUMO | $A_g$ | 0 |
| 45 | $A_g$ | $A_u$ | $A_g$ | $A_u$ | HOMO→LUMO, H-1→L+1 | $A_u$ | 0.02 |
| 60 | $A_g$ | $A_u$ | $A_u$ | $A_g$ | HOMO→LUMO | $A_g$ | 0 |
| 90 | $A_g$ | $A_u$ | $A_u$ | $A_g$ | HOMO→LUMO | $A_g$ | 0 |

In Kasha's model [1,4] (FIG. 1a), the transition dipole moment (TDM) corresponding to the $S_0$→$S_1$ transition in a monomer is described as $\mu$. $\vec{R}$ is the displacement vector connecting two molecular mass centers, θ is the angle between the vectors μ. Kasha identified J or H aggregates by the angle $\theta$. Based on the point-dipole approximation and two identical $\mu$ dipoles, when θ is smaller than 54.7 degrees, the aggregate is known as J-aggregate with the super-radiance emission. When θ is larger than 54.7 degrees, the aggregates transform into H-aggregates and the emission is quenched. To examine the relationship between $\theta$ and $f(S_1)$ by taking molecular symmetry into account, we changed θ between the two monomers of **1** from 90° to 22° (90° and 60° as the H aggregate, and 45° and 22° as the J aggregate) while keeping the centrosymmetry of $[1]_2$. TABLE. I and FIG. S5-S7 show the change in the symmetry of the dimer's MO when changing the $\theta$ value. As described by the Kasha's model, super-radiance occurs in molecular J aggregates with greater oscillation strength compared to monomeric analogues. The calculated $f(S_1)$ of $[1]_2$-J at $\theta = 22°$ is 0 by TDDFT calculations, which contrasts with Kasha's molecular exciton model. That is, regardless of the value of $\theta$, the optical selection rule constrained by crystal symmetry predicts forbidden electronic transitions in centrosymmetric J-aggregates.. Notably, To elaborate on the differences between our model and Kasha's exciton model, it should be noted that Kasha focuses on examining the interaction between two dipoles from two monomers. In contrast, our model calculates the transition dipole of the entire molecular aggregate as a whole.

For all 25 centrosymmetric dimeric compounds discussed in this work, TDDFT calculations reveal that $f(S_1)$ equals 0 in 18 molecules (**1-2**, **S1-S2**, **S8-S12**, **S14-S15**, **S16-S18**, **S20-S22**, **S24-S25**) based on their centrosymmetric packing forms and the $A_g$ symmetry of their respective $S_1$ states (TABLE SIII). Notably, six of these 18 molecules are J-aggregates. Forbidden $S_1$→$S_0$ transitions in **1-2**, **S1-S2**, **S8-S12**, **S14-S15**, **S16-S18**, **S20-S22** and **S24-S25** dimers plays an important role in the induction of UOP. When the radiative transition from the $S_1$ state to the $S_0$ state is symmetry forbidden, the non-radiative ISC process occurs efficiently, leading to the population of triplet excitons in the molecular aggregate [1]. Next, we take the dimers $[1]_2$ and $[2]_2$ as examples to consider the role of centrosymmetry in the spin-forbidden $T_1$→$S_0$ electronic transition. We used the P-SOC (perturbative SOC) method to calculate the oscillator strength of the $T_1$→$S_0$ transition in **1**, $[1]_2$, **2** and $[2]_2$. Although the calculated $f(T_1)$ values for monomers **1** and **2** are non-zero ($f(T_1) = 5 \times 10^{-6}$ in **1** and $1 \times 10^{-10}$ in **2**), the $T_1$→$S_0$ transitions of dimer $[1]_2$ and $[2]_2$ are totally forbidden ($f(T_1) = 0$). To understand the forbidden $T_1$→$S_0$ transition and the role of centrosymmetry, we calculated several parameters that determine the oscillator strength of the $T_1$→$S_0$ transition, including the matrix of $<S_m|\mu_e|S_0>$ and $<T_1|H_{SOC}|S_m>$ (m = 1-10). The TDM of the $T_1$→$S_0$ transition, $\boldsymbol{M_T}$, can be described by Equation (1). The SOC matrix describes the coupling between the $S_m$ and $T_1$ excited states, and is important for determining the oscillator strength of the $T_1$ state. By mixing the $T_1$ excited state with singlet excited states through SOC, the spin forbidden $T_1$→$S_0$ transition can be relaxed.

$$\boldsymbol{M_T} = \sum_m \frac{<T_1|H_{SOC}|S_m>}{E(S_m)-E(T_1)} \boldsymbol{M_{S_M}} \quad (1)$$

$E(S_m)$ and $E(T_1)$ are the energies of the mth singlet excited state ($S_m$) and the lowest triplet excited state ($T_1$), respectively. $\boldsymbol{M_{S_M}}$ is the TDM of the $S_m$→$S_0$ transition. Since the SOC operator is totally symmetric [18], the SOC bra-ket of $<T_1|H_{SOC}|S_m>$ is nonzero only when $T_1$ and $S_m$ contain the same symmetry ($A_g$ or $A_u$). For dimer $[1]_2$ and $[2]_2$, their $T_1$ states are derived from the LUMO→HOMO transition, where HOMO and LUMO belong to the $A_u$ symmetry. Therefore, the symmetry of the $T_1$ state in $[1]_2$ and $[2]_2$ is $A_g$. As shown in TABLE II and SIV, we calculated $f(S_m)$ (m=1-10) for $[1]_2$ and $[2]_2$ and observed nonzero values for $f(S_m)$ when $S_m$ belongs to $A_u$ symmetry. However, when $S_m$ has $A_u$ symmetry, its SOC with $T_1$ state ($T_1$: $A_g$) is zero. The calculation results in TABLE II and SIV show that if the $T_1$ state has $A_g$ symmetry, its TDM would be zero by calculating the product of $\boldsymbol{M}_{S_M}$ and $<T_1|H_{SOC}|S_m>$.

TABLE II. Calculated symmetry, $<T_1|H_{SOC}|S_m>$ and oscillator strengths (f) of $S_m$ (m=1-10) for $[\mathbf{1}]_2$. $T_1$ of $[\mathbf{1}]_2$ belongs to the $A_g$ subgroup. $<T_1|H_{SOC}|S_m>$ is calculated as the sum of squares of the SO coupling matrix of all sublevels of the uncoupled states in cm$^{-1}$.

| $[\mathbf{1}]_2$ | $S_1$ | $S_2$ | $S_3$ | $S_4$ | $S_5$ | $S_6$ | $S_7$ | $S_8$ | $S_9$ | $S_{10}$ |
|---|---|---|---|---|---|---|---|---|---|---|
| Symmetry | $A_g$ | $A_u$ | $A_g$ | $A_u$ | $A_g$ | $A_u$ | $A_u$ | $A_g$ | $A_u$ | $A_g$ |
| $<T_1|H_{SOC}|S_m>$ (cm$^{-1}$) | 0.37 | 0 | 2.59 | 0 | 3.43 | 0 | 0 | 1.81 | 0 | 4.39 |
| f($S_m$) | 0 | 0.0049 | 0 | 0.70 | 0 | 0.27 | 0.48 | 0 | 0.42 | 0 |
| f($S_m$)•$<T_1|H_{SOC}|S_m>$ | 0 | 0 | 0 | 0 | 0 | 0 | 0 | 0 | 0 | 0 |

Examination of all 25 centrosymmetric dimeric compounds reveals that 18 of them (**1-2**, **S1-S2**, **S8-S12**, **S14-S15**, **S16-S18**, **S20-S22**, **S24-S25**) have a totally symmetric $T_1$ state, which indicates a symmetry-forbidden $T_1{\rightarrow}S_0$ electronic transitions. The forbidden $T_1{\rightarrow}S_0$ electronic transition leads to small $k_r$ values of these molecular aggregates, which helps generating UOP with the long emission lifetime. The centrosymmetric packing form is closely related to UOP properties. Here some insights into the design principles of UOP materials are provided to tune the intermolecular packing form (preferred centrosymmetric) by changing the electro-negativity of the substituents. Specifically, monosubstituted benzene dimers with the mirror- or inversion-symmetry packing form (Ph-R-σ or Ph-R-*i* in FIG. 4) were used as model compounds and electro-withdrawing (-CN, -$NO_2$) and -donating (-$NH_2$) substituents are considered here. It is worth nothing that our calculations only involve the molecular dimer, whereas the prediction of crystal structure is complex and beyond the scope of this paper [19]. The intermolecular interaction energy is calculated as $E$(Ph-R-σ or Ph-R-*i*)=$E$(dimer) - $2E$(monomer). As shown in FIG. 4, compared with the mirror-symmetric dimers, all centrosymmetric dimers have stronger total intermolecular interactions, in which the electrostatic interaction plays a major role. The value of $\Delta E_{Elst}$ increases with the dipole moment of the monomer (FIG. S8-S9). We propose that the dipole-dipole interaction drives anti-parallel centrosymmetric organization, where an increase in monomer's polarity contributes to increasing the value of $\Delta E_{Elst}$ [20] and leads to a centrosymmetric packing form.

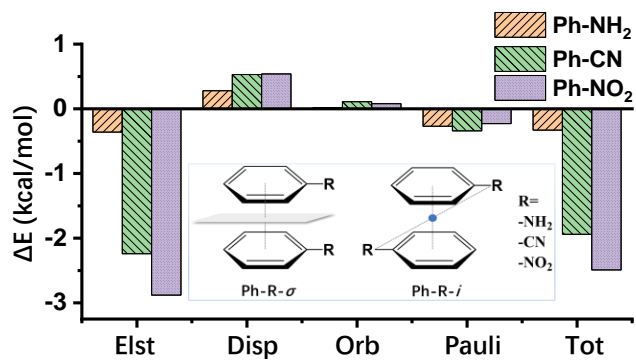

FIG. 4. Energy decomposition analysis. Elst: electrostatic interaction. Disp: dispersion interaction. Orb: orbital interaction. Pauli: Pauli repulsion. Tot: total interaction. $\Delta$ E= E(Ph-R-*i*)-E(Ph-R-σ).

To understand the photophysics of molecular aggregates, a key question recurs: do molecules emit or do not emit when they aggregate? Our study provides an efficient approach to address this question by examining the inversion symmetry of the $S_1/T_1$ excited states of aggregates and the MO interactions that ultimately determine the optical properties of the aggregates. Centrosymmetry is also closely related to UOP properties. When the excited state is totally symmetric, the centrosymmetry in the molecular aggregate leads to forbidden $S_1$ and $T_1$ electronic transitions, leading to small $k_r$ and long emission lifetime in UOP process.

### Acknowledgements

We thank the Major Program of Guangdong Basic and Applied Research (2019B030302009) and Science, Technology, and Innovation Commission of Shenzhen Municipality (JCYJ20200109150414471), the Research Grants Council (JLFS/P-704/18) of Hong Kong.

[1] M. Kasha, H. R. Rawls, and M. A. El-Bayoumi, Pure Appl. Chem. **11**, 371 (1965).
[2] Z. An *et al.*, Nat. Mater. **14**, 685 (2015).
[3] J. Mei, N. L. Leung, R. T. Kwok, J. W. Lam, and B. Z. Tang, Chem. Rev. **115**, 11718 (2015).
[4] N. J. Hestand and F. C. Spano, Chem. Rev. **118**, 7069 (2018).
[5] P. M. Kazmaier and R. Hoffmann, J. Am. Chem. Soc. **116**, 9684 (1994).
[6] J. Clark, C. Silva, R. H. Friend, and F. C. Spano, Phys. Rev. Lett. **98**, 206406 (2007).
[7] A. G. Dijkstra, H.-G. Duan, J. Knoester, K. A. Nelson, and J. Cao, J. Chem. Phys. **144**, 134310 (2016).
[8] A. Eisfeld, C. Marquardt, A. Paulheim, and M. Sokolowski, Phys. Rev. Lett. **119**, 097402 (2017).
[9] Q. Peng, Y. Yi, Z. Shuai, and J. Shao, J. Am. Chem. Soc. **129**, 9333 (2007).
[10] D. Oelkrug, A. Tompert, H.-J. Egelhaaf, M. Hanack, E. Steinhuber, M. Hohloch, H. Meier, and U. Stalmach, Synthetic metals **83**, 231 (1996).
[11] F. Würthner, Angew. Chem. Int. Ed. **59**, 14192 (2020).
[12] O. Laporte and W. F. Meggers, JOSA **11**, 459 (1925).
[13] D. S. McClure, in *Solid state physics* (Elsevier, 1959), pp. 1.
[14] L. N. Kuleshova and M. Y. Antipin, Russian chemical reviews **68**, 1 (1999).
[15] S. Hirata, Applied Physics Reviews **9**, 011304 (2022).
[16] X. F. Wang, W. J. Guo, H. Xiao, Q. Z. Yang, B. Chen, Y. Z. Chen, C. H. Tung, and L. Z. Wu, Advanced Functional Materials **30**, 1907282 (2020).
[17] S. Cai *et al.*, Advanced Functional Materials **28**, 1705045 (2018).
[18] X.-F. Wang, H. Xiao, P.-Z. Chen, Q.-Z. Yang, B. Chen, C.-H. Tung, Y.-Z. Chen, and L.-Z. Wu, J. Am. Chem. Soc. **141**, 5045 (2019).
[19] K. Saito, Y. Nakao, K. Umakoshi, and S. Sakaki, Inorg. Chem. **49**, 8977 (2010).
[20] A. Gavezzotti, Acc. Chem. Res. **27**, 309 (1994).
[21] F. Würthner, Acc. Chem. Res. **49**, 868 (2016).